\documentclass[runningheads,a4paper]{llncs}

\usepackage[intlimits]{amsmath}
\usepackage{pdfcomment}
\usepackage{amssymb}
\usepackage{graphicx}
\usepackage{braket}
\usepackage{bbold}
\usepackage{subfig}
\usepackage{floatrow}

\usepackage{booktabs}

\newcommand{\ve}[1]{\mathbf{#1}}
\newcommand{\un}[1]{\text{ #1}}

\newcommand{\md}[1]{\mathrm d #1~}

\begin{document}

\title{Ab initio Description of Optoelectronic Properties at Defective Interfaces in Solar Cells}
\titlerunning{Optoelectronic properties at interfaces from ab initio}
\author{Philippe Czaja\inst{1}
\and Massimo Celino\inst{2} \and Simone Giusepponi\inst{2}
\and Michele Gusso\inst{3} \and Urs Aeberhard\inst{1}}

\institute{IEK-5 Photovoltaik, Forschungszentrum J\"ulich, 52425 J\"ulich, Germany,\\
\email{p.czaja@fz-juelich.de}
\and ENEA, C.R. Casaccia, 00123 Rome, Italy
\and ENEA, C.R. Brindisi, 72100 Brindisi, Italy}

\authorrunning{P. Czaja et al.}

\maketitle              

\begin{abstract}
In order to optimize the optoelectronic properties of novel solar cell architectures, such as the amorphous-crystalline interface in silicon heterojunction devices, we calculate and analyze the local microscopic structure at this interface and in bulk a-Si:H, in particular with respect to the impact of material inhomogeneities. The microscopic information is used to extract macroscopic material properties, and to identify localized defect states, which govern the recombination properties encoded in quantities such as capture cross sections used in the Shockley-Read-Hall theory.
To this end, atomic configurations for a-Si:H and a-Si:H/c-Si interfaces are generated using molecular dynamics. Density functional theory calculations are then applied to these configurations in order to obtain the electronic wave functions. These are analyzed and characterized with respect to their localization and their contribution to the (local) density of states.
GW calculations are performed for the a-Si:H configuration in order to obtain a quasi-particle corrected absorption spectrum. The results suggest that the quasi-particle corrections can be approximated through a scissors shift of the Kohn-Sham energies.
\end{abstract}
\section{Introduction}
The silicon hetero-junction (SHJ) technology holds the current efficiency
record of 26.33\% for silicon-based single junction solar cells \cite{kaneka} and shows great
potential to become a future industrial standard for high-efficiency crystalline silicon (c-Si) cells.

One of the key elements of this technology is the passivation of interface defects by thin layers of hydrogenated amorphous silicon (a-Si:H), and the physical processes at the so-formed c-Si/a-Si:H interface largely influence the macroscopic characteristics of the cell.
In particular the cell performance depends critically on the optimization of transport and the minimization of recombination across the interface, which requires a profound understanding of the underlying mechanisms. Special regard has to be given to the role of localized tail and defect states in a-Si:H and at the interface, which behave substantially different from bulk states and thus prohibit a treatment in terms of bulk semiconductor physics. An accurate and physically meaningful description of the local microscopic structure is therefore an essential step in understanding and predicting the macroscopic device characteristics, which gave rise to a growing interest in ab initio approaches \cite{jarolimek:09,nolan:12,george:13}.

In our investigation presented here, we use ab initio molecular dynamics to generate atomic configurations of defective a-Si:H and c-Si/a-Si:H interfaces, and subsequently perform electronic structure calculations to obtain and characterize the electronic states.
The electronic structure at the interface is analyzed with respect to the existence of localized defect states which have an impact on the device performance due to their role as recombination centers in non-radiative recombination \cite{Shockley1952}. The density of these defect states is an important parameter in the Shockley-Read-Hall model for calculating capture cross sections, and should therefore attain realistic values in the generated structures.
The states of the bulk a-Si:H are further used for calculating the absorption coefficient from ab initio, which is a first step towards linking the global device characteristics to the local microstructure in a comprehensive multi-scale simulation approach \cite{aeberhard:16}. As the optical properties of any materials depend crucially on their band gap this quantity is of essential importance for obtaining physically relevant results. Unfortunately the independent-particle approximation, which is at the heart of standard first-principles methods, is unable to correctly predict its value \cite{perdew:85}, which is why so-called quasi-particle corrections \cite{hybertsen:85} need to be applied. The exact calculation of these corrections is however computationally expensive, a heuristic approach -- termed \emph{scissors shift} (SS) \cite{godby:88} -- , where the electron energies are simply shifted to fit the experimental band gap, is therefore often favored. Since a distinct experimental value of the band gap of a-Si:H does however not exist, a set of shifting parameters can only be determined from a quasi-particle calculation. In this paper we present the results of such a calculation for an a-Si:H configuration.

\section{Method}
\subsection{Atomic structure calculations}
The ab initio PWscf (Plane-Wave Self-Consistent Field) code of the Quantum ESPRESSO suite is used \cite{giannozzi:09,qe} to perform Born-Oppenheimer Molecular Dynamics (BOMD) simulations of the a-Si:H and the a-Si:H/c-Si interface. PWscf performs many different kinds of self-consistent calculations of electronic structure properties within Density-Functional Theory (DFT) \cite{hohenberg:64,kohn:65}, using a plane-wave (PW) basis set and pseudopotentials (PP). We use the Si and H ultrasoft pseudopotentials with Perdew-Burke-Ernzerhof (PBE) \cite{pbe} approximant 
GGA exchange-correlation potential, available in the Quantum ESPRESSO library \cite{qe}. 
To mimic infinitely extended systems, a supercell approach with periodic boundary conditions (PBC) is used.

\begin{figure}
\begin{center}
\includegraphics[width=0.27\textwidth]{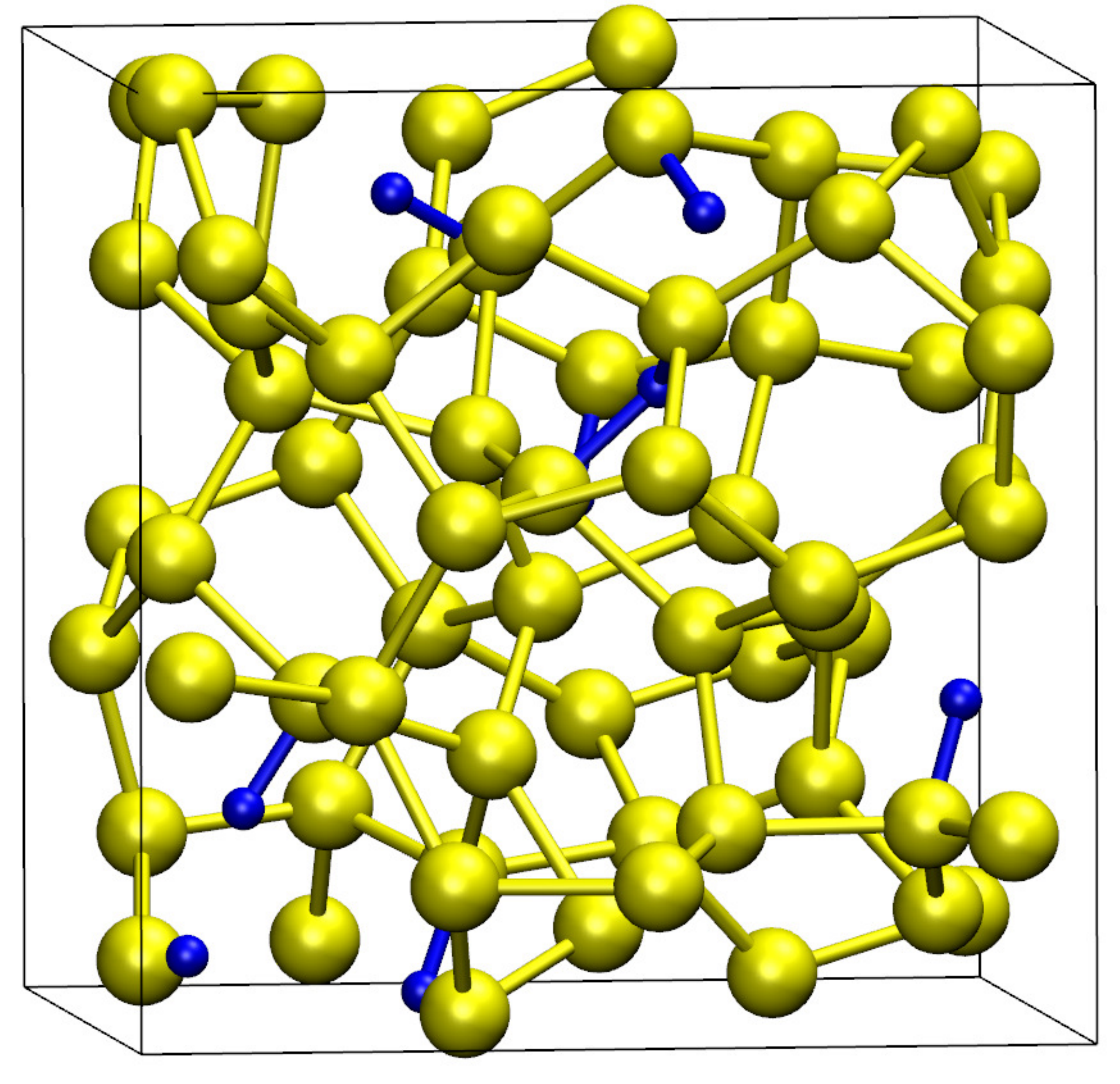}
\caption{Snapshot of the a-Si:H in the simulation box. 
Hydrogen atoms and bonds with Silicon atoms are blue, 
Silicon atoms and their bonds are yellow.}
\label{amorfo_72_finale} 
\end{center}
\end{figure}

To generate an a-Si:H system, a random starting configuration is produced with a 
percentage of H atoms of about 11\%, which is the 
nominal concentration set in experimental materials optimized for PV performance
\cite{johlin:13}.
Initially, a small system of 64 Si + 8 H atoms in a cubic supercell with size L= 11.06 \AA\ (the volume is chosen to fix the density to the experimental value of 
2.214 g/cm$^3$ \cite{curioni:11}) is used to perform calculations with a wide range of quench rates.
This is due to the fact that the resulting amorphous configuration is 
largely dependent on the quench rate used to produce the amorphous structure 
from the melt configuration.
Experimental results indicate that the amorphous phase contains a very low number of defects and that the majority of Si atoms have 
coordination four.
To this end we select a small amorphous configuration (Fig. \ref{amorfo_72_finale}) that 
minimizes both the total value of defects and the deviation from the fourfold 
coordination of the Si atoms. Then, this configuration, is used as starting configuration for a BOMD simulation on the electronic ground state at constant volume and constant temperature for 6.5 ps,
controlling the ionic temperature (T = 300 K) by using an Andersen thermostat \cite{andersen:80}.

The final configuration is then used to produce a large system by
replicating it in all directions.
The resulting large system is thus composed of 
512 Si + 64 H atoms and has a size of L=22.12 \AA. Due to the high computational costs required by PWscf, BOMD simulations on this large system are performed with the Quickstep code of the CP2K suite \cite{cp2k}.
CP2K is a quantum chemistry and solid state physics software package
that can perform atomistic simulations with different modelling methods
(such as DFT) using a mixed Gaussian and plane wave approach.
Norm conserving Goedecker-Tetter-Hutter
pseudopotentials with PBE exchange-correlation and an optimized TZV2P gaussian basis set are used \cite{goedecker:96,hartwigsen:98,krack:05}.
Self consistency at each MD step is achieved using the orbital transformation method \cite{vandevondele:03}.
An annealing process from
T = 300 K up to T = 600 K, and then back to T = 300 K for 60 ps was then
used to thermalize the whole atomic configuration, and minimize the defects at
the internal interfaces. After the annealing, a simulation run at T = 300 K was
performed for about 20 ps.

The a-Si:H/c-Si interface is built by putting nearby two free surfaces obtained cutting both the crystalline silicon and the hydrogenated amorphous silicon. The relaxed p($2 \times 1$) symmetric reconstruction of the Si(001) surface constitutes the c-Si side of the interface.
It is formed by 192 Si atoms: 12 layers of silicon each of them with 16 atoms. 
The a-Si:H side of the system is generated using a simulated-annealing quench-from-the-melt simulation protocol and is composed of 128 Si atoms and 16 H atoms. A void region of about 10~\AA is added to suppress the interaction between the external surfaces due to PBC. This distance was checked by convergence tests.
The total length of the system is L$_z$ =38.70 ~\AA, while in the x and y direction the system has L$_x$ = L$_y$ = 15.48~\AA.
Total energy calculations of the system at different distances between c-Si and a-Si:H, were performed to find the interface configuration corresponding to the lowest total energy. The configurations were built moving rigidly by hand the a-Si:H part and keeping fixed the c-Si one.

\floatsetup[figure]{style=plain,subcapbesideposition=top}
\begin{figure}[b!]
\begin{center}
\sidesubfloat[]{
\includegraphics[width=0.7\textwidth]{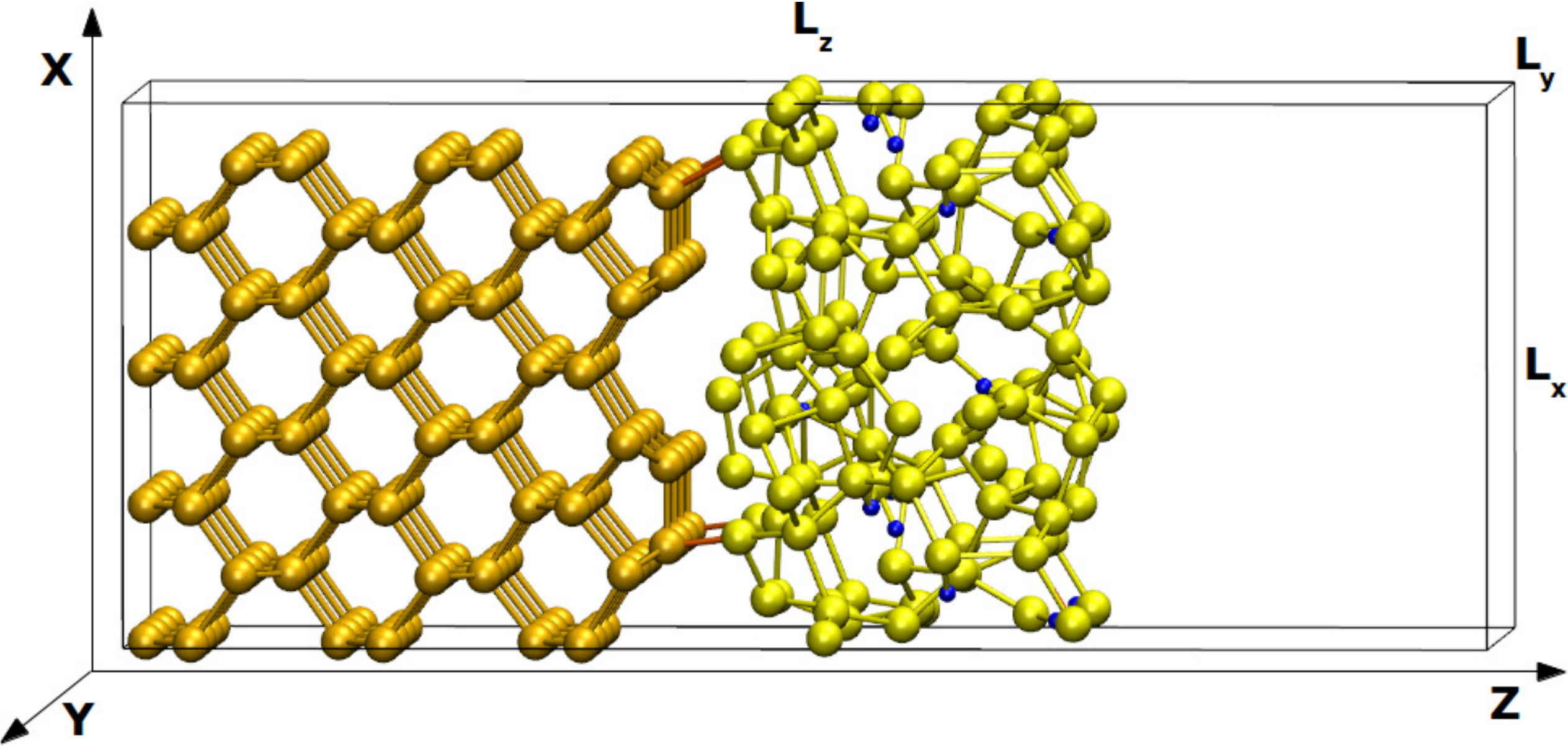}
\label{conf_ini} 
}

\sidesubfloat[]{
\includegraphics[width=0.7\textwidth]{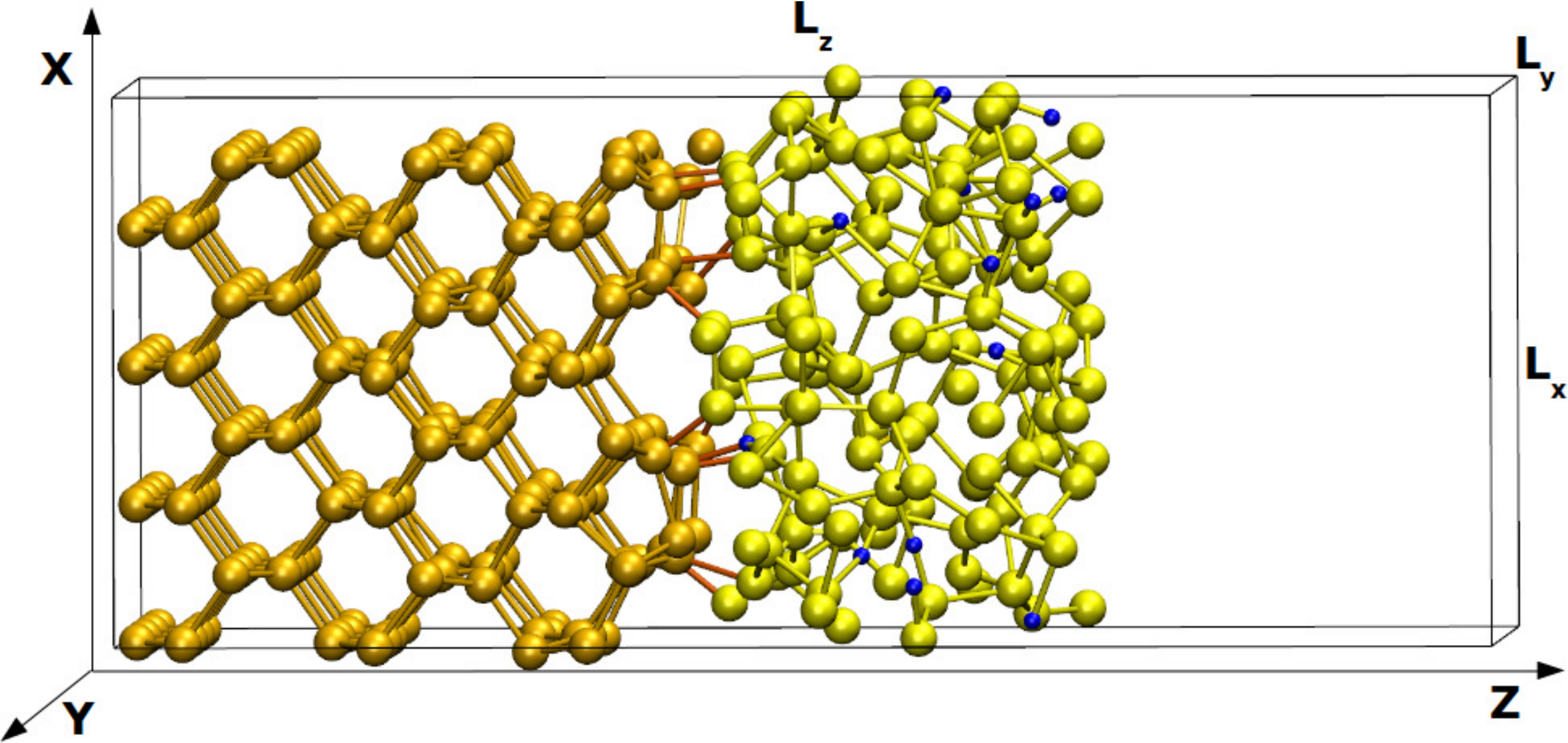}
\label{conf_fin} 
}
\caption{Snapshots of the a-Si:H/c-Si interface in the simulation box. 
The structure is infinitely extended in both x and y directions. 
A void region is considered to suppress the interaction between the external surfaces
due to periodic boundary conditions.
Free surfaces and a-Si:H/c-Si interface are perpendicular to the y axis. 
Hydrogen atoms are blue, silicon atoms are dark yellow in the c-Si part and light yellow in the a-Si:H part. (a) Initial configuration. (b) Configuration at 35 ps of the MD simulation. The Si atoms near the interface have moved to form bonds between the c-Si and the a-Si:H layer.
}
\end{center}
\end{figure}

The interface shown in Fig. \ref{conf_ini}, is used as starting configuration for MD simulation on the electronic ground state at constant volume and constant temperature (NVT). The ionic temperature is fixed at T = 300 K and is controlled using an Andersen thermostat \cite{andersen:80}. The first four layers of c-Si atoms on the left are kept fixed to impose a bulk like behavior to the crystalline silicon part of the system. The MD simulation is performed for more than 35~ps, the initial part of the simulation (20~ps) was used to thermalize the system and reach a stable configuration. Figure \ref{conf_fin} displays the configuration of the a-Si:H/c-Si interface at 35~ps.

\subsection{Electronic structure calculations}
We use density functional theory (DFT) \cite{hohenberg:64,kohn:65} with periodic boundary conditions to self-consistently calculate the electronic structure of the a-Si:H and the interface configurations described above. The interface configuration is enclosed in a super cell that includes an additional vacuum layer to avoid self-interaction. All calculations are done with the PW-PP code Quantum ESPRESSO \cite{giannozzi:09,qe} using the PBE-GGA exchange-correlation functional \cite{pbe}. For the c-Si/a-Si:H interface a k-point grid of size 4x4x1 and a plane-wave cut-off of 28 Ry is used, for the a-Si:H a 4x4x4 (72 atom configuration) and a 2x2x2 grid (576 atom configuration) respectively, together with a cut-off energy of 52 Ry.

Subsequent to the electronic structure calculation the wave functions and electronic density of states (DOS) of the c-Si/a-Si:H interface are analyzed to obtain information about its local microscopic properties, which are relevant for the mesoscopic dynamics and macroscopic device characteristics. In particular the wave function localization is analyzed qualitatively and quantitatively to allow for the distinction of localized states and the identification of their origins. In combination with the local DOS the contribution of dangling bonds and interfaces to the important mid-gap states can be determined.

As a quantitative measure for the localization of the wave function $\psi$ we use the spread $S$, which is calculated as the square root of the variance of $|\psi|^2$ with respect to the super cell:
\begin{equation*}
  S_z = \sqrt{12\left(\braket{z^2} - \braket{z}^2\right)} = \sqrt{12}\cdot\sqrt{\int_{\Omega}\md{\ve r} |\psi(\ve r)|^2 z^2 - \left(\int_{\Omega} \md{\ve r} |\psi(\ve r)|^2 z \right)^2} \text{ ,}
\end{equation*}
where we assume that $\psi$ is normalized. It can be easily seen that a maximally localized $\psi$ (i.e., a delta function) gives $S_z = 0$, whereas a wave function that is maximally delocalized over the super cell (i.e., a plane wave) will result in $S_z = L$, where $L$ is the length of the super cell. This explains the factor $\sqrt{12}$ in the definition.
The integration volume $\Omega$ is naturally chosen such that the boundaries lie inside the vacuum layer where $\psi\approx 0$, such that shifting the integration volume does not affect $S$. This also provides an unambiguous definition of the wave function center
\begin{equation*}
\braket{z} = \int_{\Omega} \md{\ve r} |\psi(\ve r)|^2 z \text{ ,}
\end{equation*}
which can be interpreted as the position where the wave function is localized.
This definition allows us to identify localized states (i.e., states with small spread), and to locate them both in real and in energy space.

In order to relate the electronic properties of the interface to the atomic structure, and in particular investigate the effect of structural defects, we use the electron localization function (ELF) \cite{becke:90}, which enables us to determine the coordination of each atom, and to identify dangling bonds and weakly bond atoms. For that purpose the ELF is computed along the axes between neighboring atoms, where it shows a characteristic behavior for covalent bonds \cite{savin:92}. This is performed with the Quantum ESPRESSO package.

\subsection{Optical calculations}
\begin{figure}
\begin{center}
\includegraphics[width=0.52\textwidth]{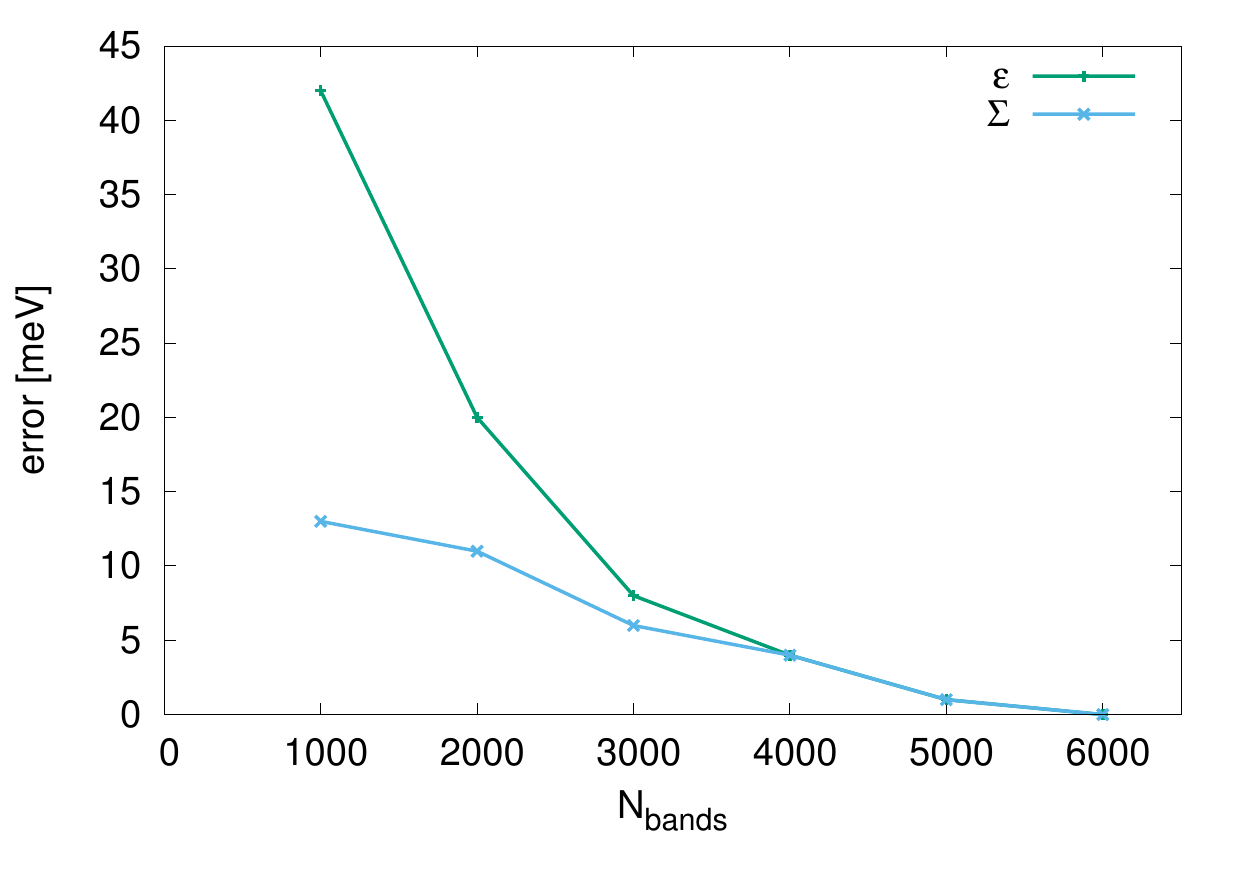}
\caption{Convergence of band gap with respect to the number of bands included in the calculation of $\epsilon$ and $\Sigma$ respectively.}
\label{convergence} 
\end{center}
\end{figure}

The calculation of the absorption coefficient for the 72-atom a-Si:H configuration is carried out within the random phase approximation (RPA) \cite{ehrenreich:65} as implemented in the BerkeleyGW code \cite{Deslippe2012}, using the non-interacting Kohn-Sham states on a 2x2x2 k-point grid. The same code is used for calculating the quasi-particle (QP) corrections to the Kohn-Sham energies with the GW formalism \cite{Hedin1965}. In order to reduce the computational costs we perform a single-shot G$_0$W$_0$ calculation together with the plasmon-pole approximation \cite{lundqvist:67,overhauser:71,hybertsen:86}. This has the advantage of requiring the dielectric tensor $\epsilon(\omega)$ only in the static limit $\omega\to 0$, as opposed to a full-frequency calculation, while offering similar accuracy for many semiconductors, including c-Si \cite{larson:13}. The band gap is converged with respect to the cut-off energy $E_{cut}^{\epsilon}$ used in the calculation of $\epsilon$, for which we find a value of 10 Ry, and with respect to the number of unoccupied bands $N_{\mathrm{bands}}^{\epsilon}$ and $N_{\mathrm{bands}}^{\Sigma}$ included in the calculation of $\epsilon$ and of the self energy $\Sigma$. We find that a large number of roughly 3000 bands is needed to reach convergence of both quantities (Fig. \ref{convergence}). The absorption coefficient is recalculated with the corrected energies $E^{QP}$, and compared to calculations where scissors shifts with different sets of parameters are used. These parameters are obtained by applying a linear fit $E_{v/c}^{QP} = a_{v/c}\cdot E_{v/c}+b_{v/c}$, where $E_{v/c}$ are the uncorrected energies, both to the valence and the conduction band. The absorption calculation for the 576-atom configuration is carried out on a 2x2x2 k-point grid as well, using uncorrected and scissors shift energies.

\section{Results}

\subsection{a-Si:H}
\begin{figure}
\begin{center}
\includegraphics[width=0.52\textwidth]{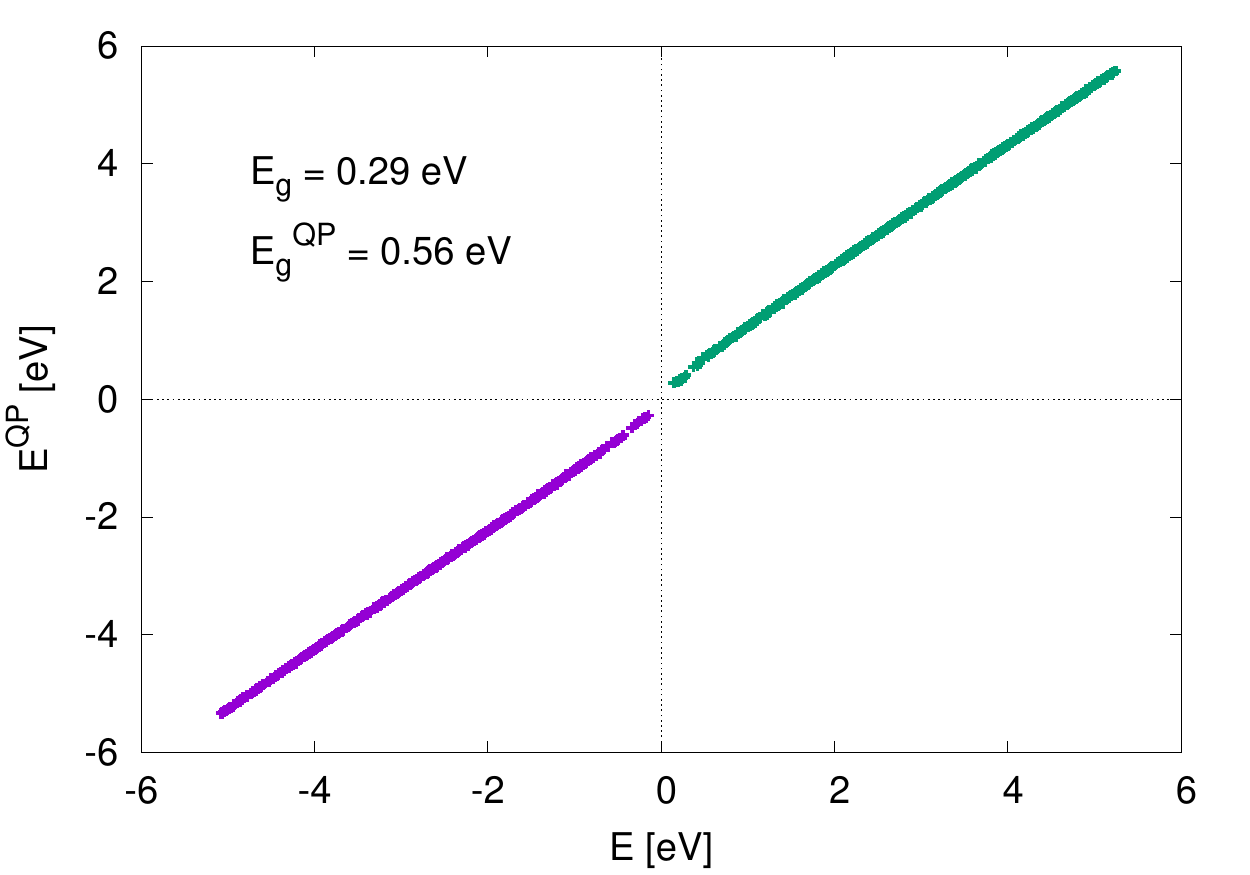}
\caption{Quasi-particle corrected vs. uncorrected electron energies. $E_g$ refers here to the energy difference between the lowest unoccupied and the highest occupied state.}
\label{qp} 
\end{center}
\end{figure}

Figure \ref{qp} shows the quasi-particle corrected electron energies as obtained from the GW calculation for the 72-atom a-Si:H structure described above. The results show that the effect of the corrections consists mainly in a spreading of valence and conduction band by approximately $0.26$ eV. This suggests that the costly GW calculation can be substituted by a simple scissors shift in further calculations of a-Si:H structures. The choice of the right set of parameters depends on the energy range of interest. By applying a linear fit in the energy range from $-1$ to $1$ eV we obtain $a_v = 1.088$, $b_v = -1.097$ eV, $a_c = 1.146$, and $b_c = -1.228$ eV.

\begin{figure}[t!]
\begin{center}
\includegraphics[width=0.5\textwidth]{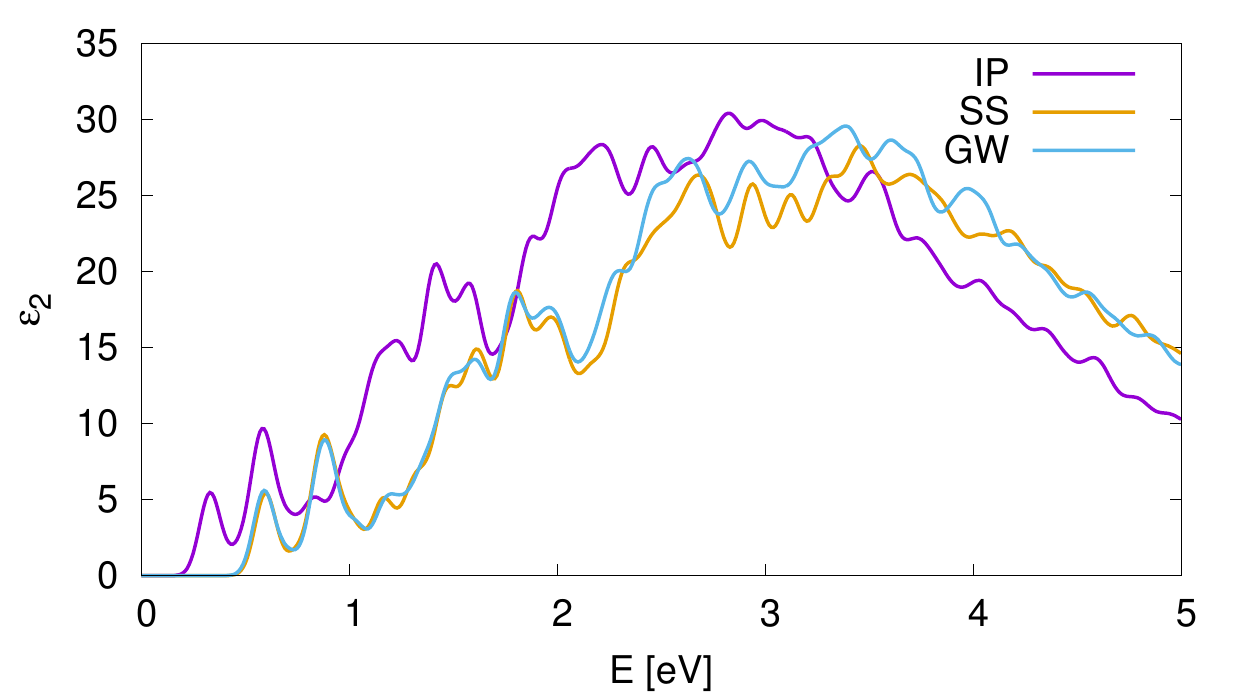}\includegraphics[width=0.5\textwidth]{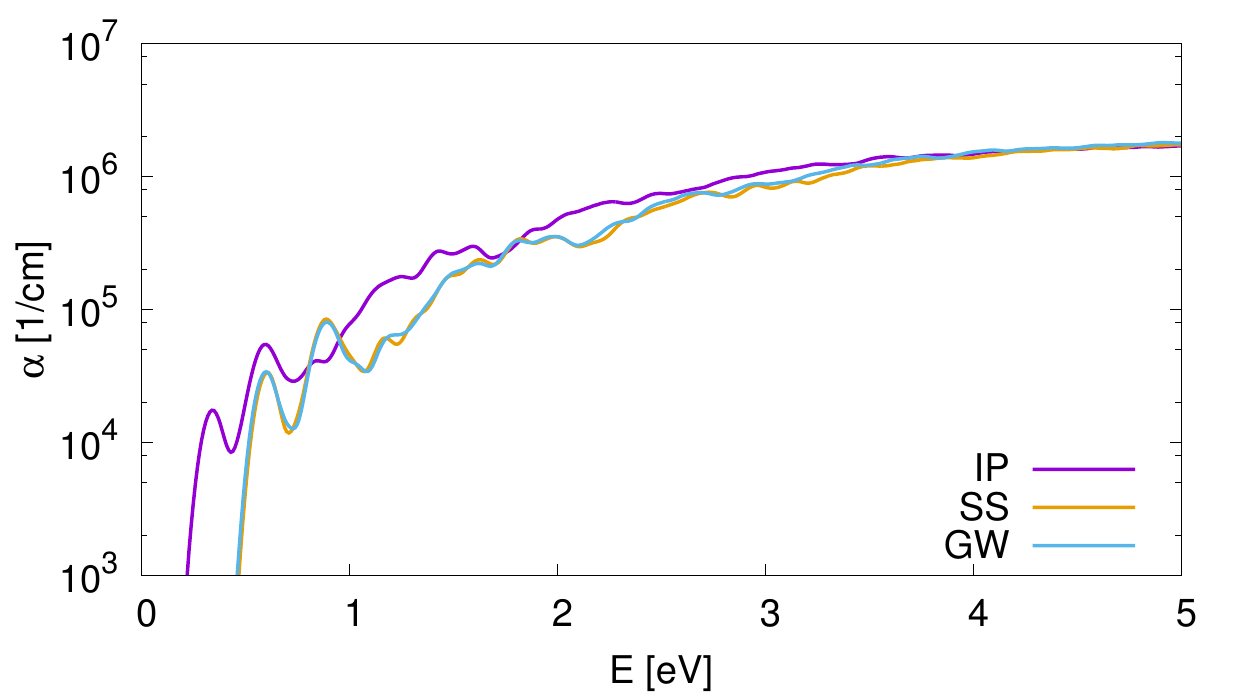}
\caption{Imaginary part of dielectric function (left) and absorption coefficient (right) for the 72-atom configuration, calculated with uncorrected states (IP), and with quasiparticle corrected states in GW and scissors shift (SS) approximation.}
\label{alpha} 
\end{center}
\end{figure}

Figure \ref{alpha} shows the imaginary part of the dielectric function and the absorption coefficient calculated within the independent-particle approximation, that is, with the uncorrected Kohn-Sham energies, the GW approximation, and the scissors-shift approximation. The GW correction modifies the absorption spectrum only in terms of a shift and a slight stretch. This correction can be very well approximated by a scissors shift with the parameters given above, which reproduce almost exactly the GW absorption spectrum.

\begin{figure}[t!]
\begin{center}
\includegraphics[width=0.5\textwidth]{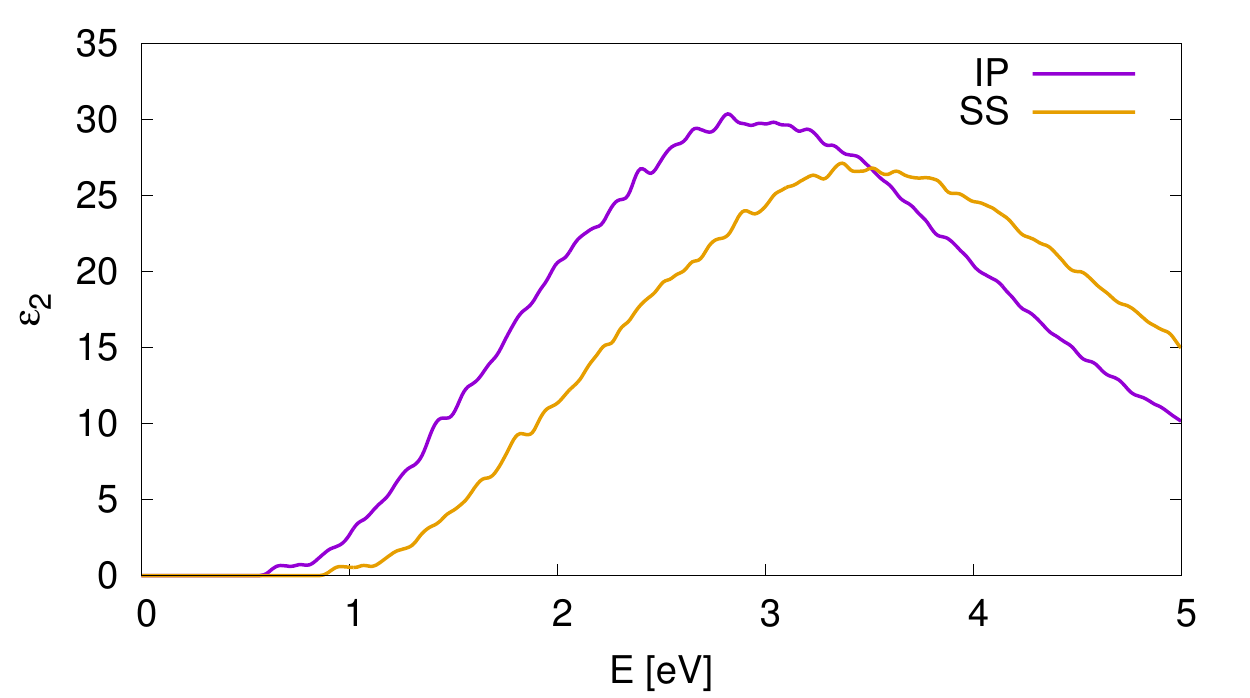}\includegraphics[width=0.5\textwidth]{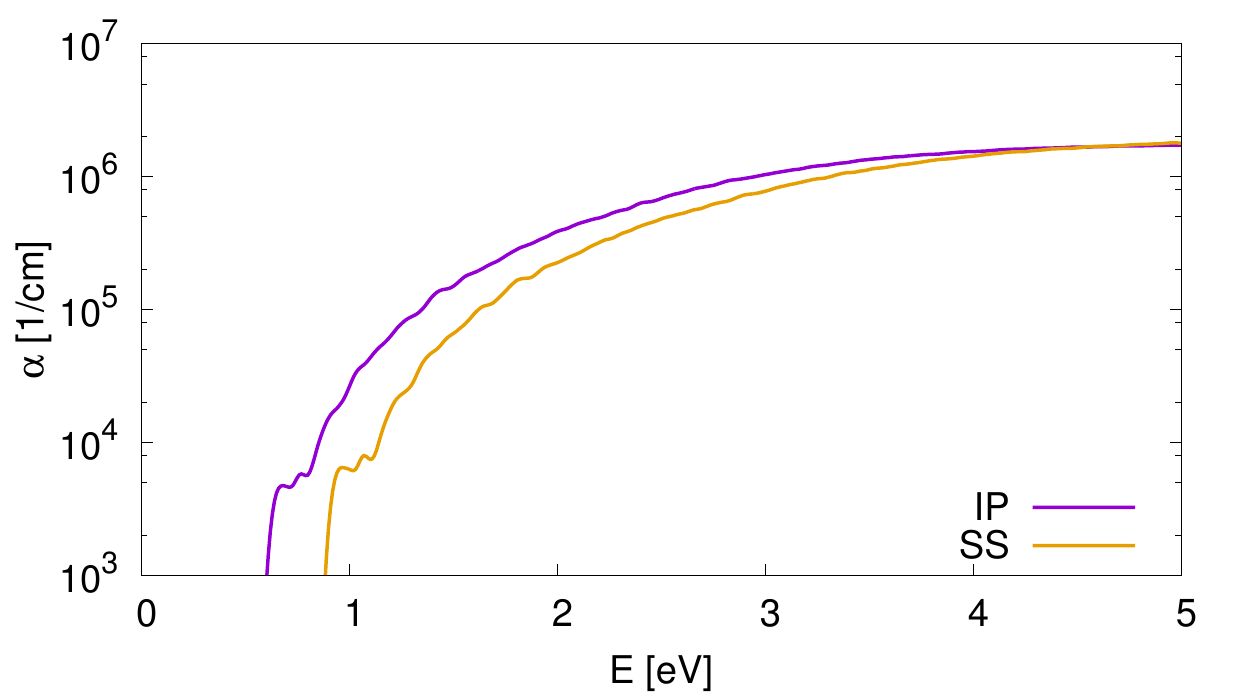}
\caption{Imaginary part of dielectric function (left) and absorption coefficient (right) for the 576-atom configuration, calculated with uncorrected states (IP), and with quasiparticle corrected states in scissors shift (SS) approximation.}
\label{alpha512} 
\end{center}
\end{figure}

Using the scissors shift approximation enables us to calculate a quasi-particle-corrected absorption spectrum also for the 576-atom structure, which is shown in Fig. \ref{alpha512}. Comparison of the spectra for the two different configurations shows an increase of the optical band gap in the larger structure, along with a decrease of the sup-gap absorption peaks. Even though this represents an improvement, the band gap is still small compared to the experimental value of $1.7$ eV, which suggests that one might have to go to even larger structures in order to eliminate finite-size effects that artificially reduce the band gap, like the overestimation of the defect density and interactions with periodic images.


\subsection{c-Si/a-Si:H interface}

\begin{figure}
\begin{center}
\includegraphics[width=0.65\textwidth]{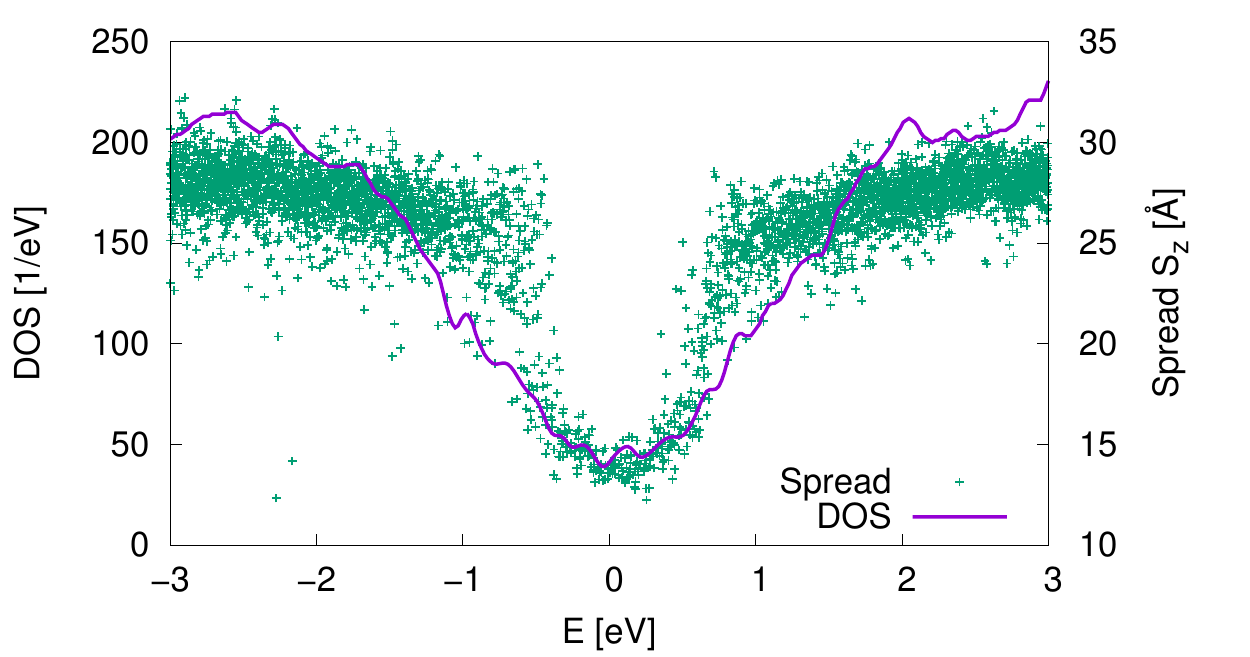}
\caption{DOS and wave function spread (in direction perpendicular to the interface) in the energy region around the Fermi energy at 0 eV.}
\label{DOS} 
\end{center}
\end{figure}

In Fig. \ref{DOS} the spread in z-direction (i.e., in growth direction)
$S_z$ is shown as a function of the wave function energy together with
the total DOS around the Fermi energy for the above described
interface configuration. The figure shows that there is a dense distribution
of strongly localized states inside the c-Si band gap, which can be clearly
distinguished from the more extended tail and bulk states.

\begin{figure}[t!]
\begin{center}
\includegraphics[width=0.62\textwidth]{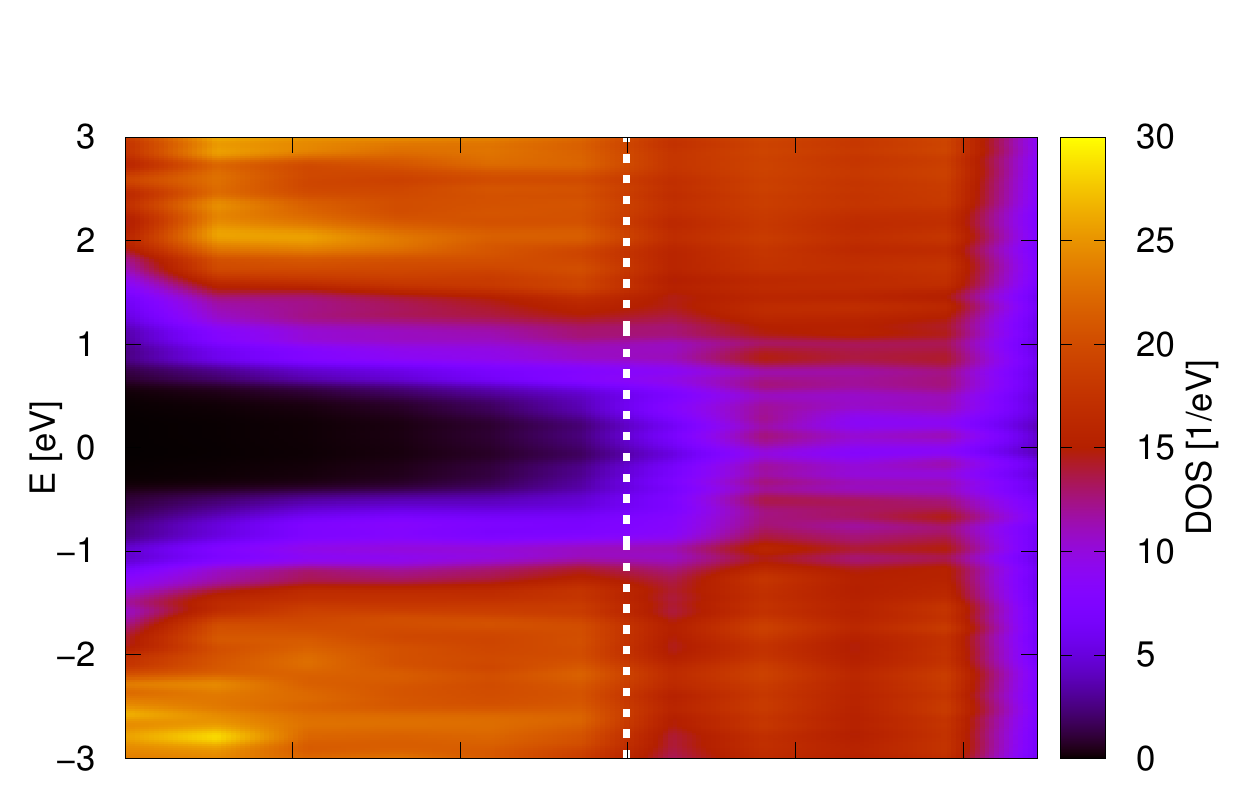}
\includegraphics[width=0.62\textwidth]{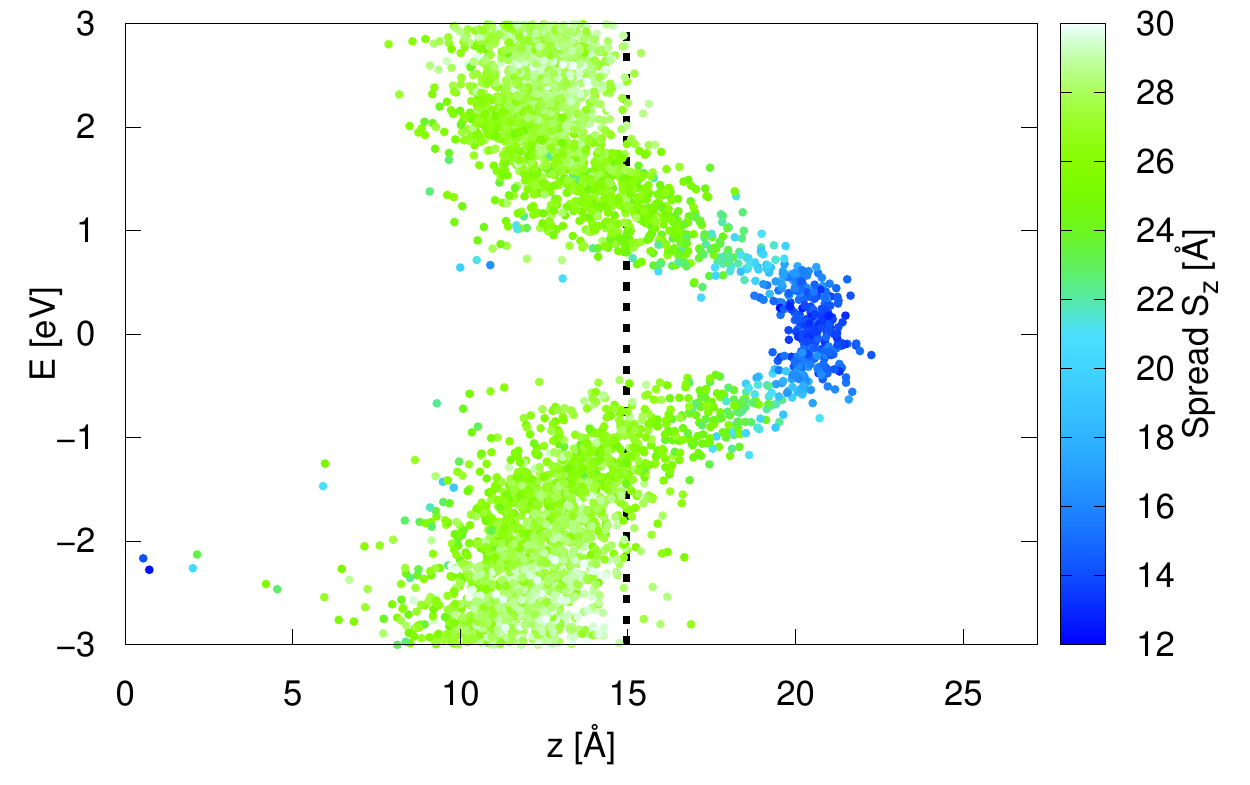}
\caption{Top: Local DOS integrated over layers parallel to the interface as a function of the z-coordinate. Bottom: Wave function spread in z-direction. Each dot marks the energy and the position of the center of one wave function, whereas the color represents its spread. The dotted line shows the approximate position of the interface.}
\label{spread} 
\end{center}
\end{figure}

The origin of these
states can be investigated further by looking at the local DOS and the wave
function centers as shown in Fig. \ref{spread}. In the top subfigure, the
layer-resolved DOS is displayed as a function of the z-coordinate, which is
obtained by integrating the local DOS over layers parallel to the interface. The figure shows that near the interface, which is marked by the dotted line, the band gap starts filling up with states, and completely vanishes in the a-Si:H region. That these mid-gap states are indeed localized can be seen in the bottom figure, where $S_z$ is plotted as a function of the energy and the z-component of the
center of the wave functions. Each dot marks the energy and the position of
one wave function, that is, where along the z-direction it is centered. The color of
each dot represents the spread of the wave function. This representation indicates that the contribution to the mid-gap states comes mainly from localized states in
the a-Si:H layer, whereas the interface region hardly contributes at all.

\begin{figure}[t!]
\begin{center}
\includegraphics[width=0.9\textwidth]{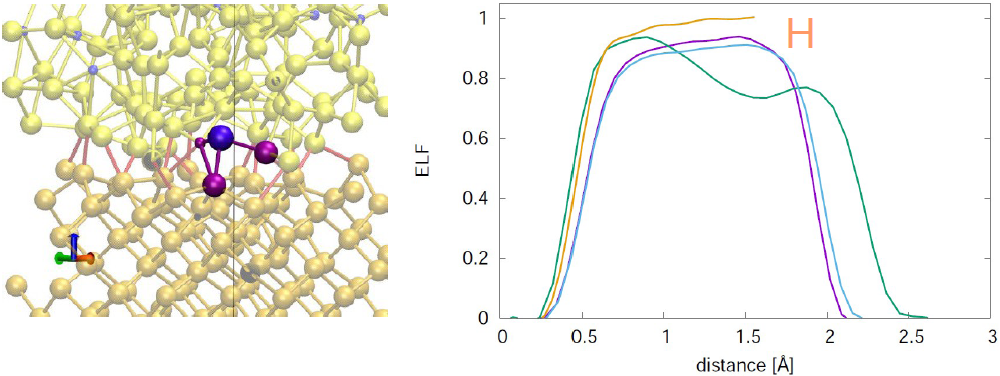}
\caption{Left: 3-fold bond atom at the a-Si:H/c-Si interface (purple) and its three bonding partners (magenta). Right: ELF between the atom shown on the left and its four nearest neighbors. The orange curve represents a bond with an H atom, the blue and purple curve represent Si-Si bonds, and the green curve represents a dangling bond.}
\label{elf} 
\end{center}
\end{figure}

\begin{figure}[t!]
\begin{center}
\includegraphics[width=0.84\textwidth]{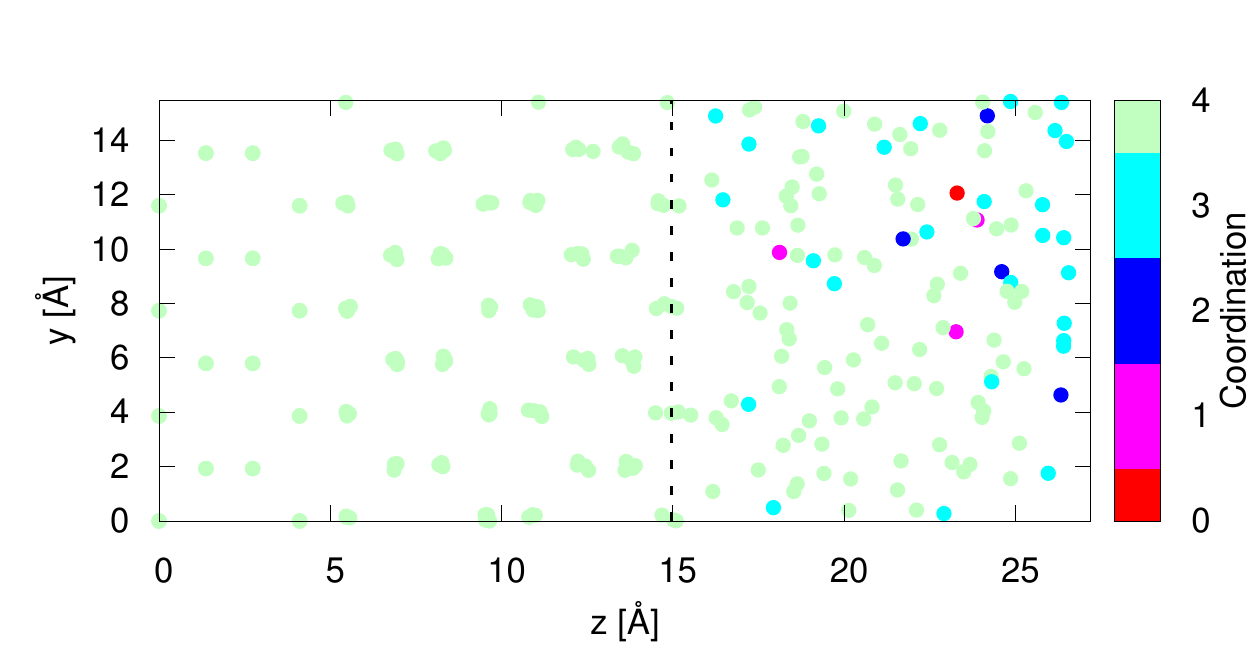}
\caption{Coordination numbers for all atoms in the a-Si:H/c-Si configuration.}
\label{coordination} 
\end{center}
\end{figure}

The emergence of localized states in the a-Si:H region can be better understood in terms of the atomic structure. For that purpose all the bonds are analyzed by means of the ELF in order to identify dangling and weak bonds. This is shown exemplarily in Fig. \ref{elf} for a three-fold bond Si atom. By investigating the ELF between this atom and its nearest neighbors one can clearly distinguish one H bond, two Si bonds, and one dangling Si bond. Applying this analysis to all atoms yields a coordination map as shown in Fig. \ref{coordination}. This reveals that there is a large number of low-coordinated atoms in the a-Si:H layer whereas the atoms at the interface itself (represented by a dotted line) are mostly four-fold coordinated. While supporting the conclusions from the localization analysis, this result also indicates that the quality of the amorphous layer is rather poor. In fact the defect density is of the order of $10^{22}/\un{cm}^3$, and thus four orders of magnitude higher than the defect density measured experimentally for thin a-Si:H films \cite{favre:87}, which explains the high DOS inside the band gap.


\section{Computational costs}
Table \ref{cpu_time} lists the computational costs of typical sets of calculations with converged parameters for all three structures considered in this paper. As self-consistent field (scf) calculations with DFT scale with $\mathcal O(N^2\log N)$ in the number of atoms, the computational costs increase by two orders of magnitude when going from 72 to 576 atoms. This is however only the theoretical scaling behavior and does not yet take into account I/O and communication costs, as well as the non-ideal scaling with respect to the number of cores. These effects become visible especially in the non-self consistent (nscf) calculation of unoccupied bands, where larger matrices have to be handled. Altogether the numbers indicate that the current limit for performing DFT calculations with conventional plane-wave approaches on reasonable time-scales is of the order of a few thousand atoms.

Regarding GW, a full calculation for the small structure requires about 2400 core-h, making it obvious that GW calculations for the larger systems are currently out of range. The most costly part here is the DFT calculation of a large number of unoccupied bands, which are needed exclusively in the GW calculations and not for any other of the calculations we performed. In addition it has to be pointed out once again that the numbers given here refer to an already converged calculation. The convergence process itself is computationally much more challenging, which is due to the fact that three interdependent parameters have to be converged simultaneously, resulting in a total cost of about 80000 core-h.

For the large a-Si:H and the interface configuration the vast majority of the computational time is spent on the BOMD simulations which is due to the fact that the electronic ground state is computed at every time step by an scf calculation. The high computational demand for the interface generation motivated the use of the CP2K Quickstep code for the large a-Si:H system, which allowed us to reduce the computational cost by using a mixed Gaussian and plane wave approach.

\begin{table}
\caption{Computational costs for typical sets of calculations for all three structures investigated within this work. MD and DFT calculations were done using Quantum ESPRESSO, except
for the 576-atom system, where CP2K was used. GW and absorption calculations were done using BerkeleyGW. For the MD calculations also the simulation time is provided in brackets.}
\begin{tabular}{@{}ll|rrr@{}}
\hline
\hline
\rlap{Calculation} & & \multicolumn{3}{@{}c@{}}{Computational costs [core-h]}\\
& & a-Si:H (72) & a-Si:H (576) & a-Si:H/c-Si \\ 
\hline 
\rlap{MD} &  & 2300 & 190000 & 220000 \\ 
& & (6.5 ps) & (80.0 ps) & (35.0 ps) \\
DFT & scf & 20 & 1940 & 200 \\ 
 & nscf & 15 & 5450 & 750 \\ 
GW & unoccupied bands & 1800 &  &  \\ 
 & $\epsilon$ & 380 &  &  \\ 
 & $\Sigma$ & 180 &  &  \\ 
\rlap{Absorption} & & 1 & 320 &  \\ 
\hline
\hline
\end{tabular}
\label{cpu_time}
\end{table}

\section{Conclusions}
We presented an ab initio description of the atomic and electronic properties of the a-Si:H/c-Si interface, which is at the heart of the technologically relevant silicon-heterojunction solar cells. We introduced and applied different methods for analyzing the electronic structure, in particular with respect to the role of defects and localized states, which have an influence on the cell performance via non-radiative recombination.

Furthermore, we generated configurations of a-Si:H and calculated the electronic structure, including GW corrections. As a first step towards the extraction of macroscopic material properties from the local microscopic structure for use in multiscale models of solar cells, we calculated the absorption spectrum of an a-Si:H structure. As an important result we found that the expensive GW corrections can be replaced by a linear approximation, which makes calculations for larger -- and, thus, physically more representative -- configurations possible.

\subsubsection*{Acknowledgments.}
This project has received funding from the European Commission Horizon
2020 research and innovation program under grant agreement No. 676629.
The authors gratefully acknowledge the computing time granted on the
supercomputer JURECA \cite{jureca} at J\"ulich Supercomputing Centre (JSC) and on the supercomputer CRESCO \cite{cresco} on the ENEA-GRID infrastructure.

\bibliographystyle{splncs}

\end{document}